\documentclass[12pt,a4paper]{article}
\usepackage[truedimen,margin=30mm]{geometry} 

\usepackage{mathrsfs}
\usepackage{tabularray}
\usepackage{amssymb}
\usepackage{amsmath}
\usepackage{ascmac}
\usepackage{amsthm}
\usepackage[dvips]{graphicx}
\usepackage{natbib}
\usepackage{setspace}
\usepackage{times}
\usepackage{graphicx} 
\usepackage{float} 
\usepackage{subfigure} 
\usepackage{color}
\usepackage{hyperref}  
\usepackage{amsmath}  
\usepackage{enumitem}
\usepackage{algorithm}  
\usepackage{algorithmic}
\usepackage{booktabs} 
\usepackage{array} 
\usepackage{comment}

\usepackage{titlesec}
\titleformat*{\section}{\large\bfseries}
\titleformat*{\subsection}{\it}

\newtheorem{prp}{Proposition}

\def\Wt{{\widetilde{W}}}

\title{{\bf Robust Global Fr\'echet Regression via \\
Weight Regularization}}

\date{}

\begin{document}

\maketitle
\doublespacing

\vspace{-2cm}
\begin{center}
{\large 
Hao Li$^1$, Shonosuke Sugasawa$^{2*}$ and Shota Katayama$^{2}$
}

\vspace{0.5cm}
$^1$Graduate School of Economics, Keio University\\
$^2$Faculty of Economics, Keio University\\
$^{*}$Corresponding author (Email: sugasawa@econ.keio.ac.jp)
\end{center}

\vspace{0.5cm}
\begin{center}
{\large\bf Abstract}
\end{center}

The Fr\'echet regression is a useful method for modeling random objects in a general metric space given Euclidean covariates. However, the conventional approach could be sensitive to outlying objects in the sense that the distance from the regression surface is large compared to the other objects. In this study, we develop a robust version of the global Fr\'echet regression by incorporating weight parameters into the objective function. We then introduce the Elastic net regularization, favoring a sparse vector of robust parameters to control the influence of outlying objects. We provide a computational algorithm to iteratively estimate the regression function and weight parameters, with providing a linear convergence property. We also propose the Bayesian information criterion to select the tuning parameters for regularization, which gives adaptive robustness along with observed data. The finite sample performance of the proposed method is demonstrated through numerical studies on matrix and distribution responses.

\bigskip\noindent
{\bf Key words}: 
distribution regression; network regression; random objects; robust estimation; sparsity

%---------------------------------------%
%           Introduction                %
%---------------------------------------%
\section{Introduction}
In recent years, the regression methods for response variables on manifolds have become increasingly popular, including probability distribution responses \citep{hartung2001tests}, covariance matrices \citep{newey1986simple}, network responses \citep{bar2004response}, and other complex objects. 
Their use is becoming more widespread in real-world data analysis, particularly in medicine, geological science, and logistics, as \citep{marron2014overview}. However, traditional regression techniques, which are designed for Euclidean-valued responses, are inadequate for modeling such complex data structures. To address this challenge, several recent studies have investigated regression models for non-Euclidean and manifold-valued data. 
\cite{fletcher2011geodesic} proposed a geodesic regression model called ``global Fr\'echet regression", which is the natural generalization of linear regression and is parameterized by an intercept and slope term.
\cite{miller2004computational} and \cite{jupp1987fitting} proposed an unrolling method on shape spaces. Fr\'echet regression, based on the Fr\'echet mean, has emerged as a powerful extension, enabling regression analysis when the response variable lies in a non-Euclidean metric space.

A notable limitation of the existing Fr\'echet regression is the sensitivity to outlying observations. However, research on robust regression methods in manifold spaces remains very limited. To the best of our knowledge, the only related work is that of \cite{lee2024huber} and \cite{NIPS2009_92977ae4}, who proposed and systematically analysed the Huber mean on Riemannian manifolds and provided an iterative algorithm for parameter estimation. It should be noted that each iteration of this algorithm requires geometric operations on the manifold, such as the exponential and logarithmic maps. Specifically, the logarithmic map takes each data point. It transforms it to a vector in the tangent space at the current mean, effectively describing the direction and distance from the current mean to that data point on the manifold. Since these log and exp maps do not have explicit analytical formulas for most manifolds and must be computed numerically, the computational cost of each iteration is substantially increased.

In this work, we propose a novel approach to the global robust Fr\'echet regression developed by \cite{petersen2019frechet}. The original Fr\'echet regression provides a principled approach for modeling regression relationships between vectors of real-valued predictors and complex response objects residing in a general metric space. However, similar to the aforementioned method, the standard Fr\'echet regression lacks robustness, making it sensitive to outliers and deviations. Thus, we incorporate a weight parameter (taking values on $[0,1]$) into the original objective function of the Fr\'echet regression and give the Elastic net penalty term to the weight parameter. Under this framework, observations identified as outliers are assigned weights close to zero, effectively reducing their influence on the regression estimation, whereas typical observations receive weights near one. However, direct regularization of the weight parameters themselves would undesirably shrink all weights towards zero, thereby diminishing the influence of all observations, including those that are not outliers. 
Our new methodology offers two key advantages. First, under both the Frobenius distance and the $L_2$ Wasserstein distance, it allows for closed-form solutions for the estimators, thereby facilitating efficient computation. Second, our simulation demonstrates that the proposed algorithm exhibits rapid convergence, often requiring only a small number of iterations to achieve stable estimates.
For the selection of optimal tuning parameters in the regularization term, we adopt the Bayesian information criterion (BIC), following \cite{gao2016penalized}, who proposed a weighted model for response variables in the Euclidean space.
Building upon this approach, we extend the methodology to accommodate situations where the response variables reside in non-Euclidean spaces.

The rest of the paper is organized as follows. Section 2 provides an overview of global robust Fr\'echet regression, introduces the framework of robust Fr\'echet regression with weight regularization, discusses the linear convergence properties of the proposed optimization algorithm, and describes the procedure for selecting tuning parameters using the BIC criterion. In Section 3, we demonstrate the applicability of our approach to both matrix-valued and distribution-valued responses,and present a fixed-point algorithm for implementation, and report simulation results along with analysis on real-world datasets to illustrate the effectiveness of the proposed method. In Section 4, we provide a brief discussion of the methods and possible extensions.
R code implementing the proposed method is available at the GitHub repository
(\url{https://github.com/lee199950120HAO/robust-FR}).

%---------------------------------------%
%              Methods                  %
%---------------------------------------%
\section{Robust Global Fr\'echet Regression}

\subsection{Global Fr\'echet regression}
We first briefly introduce the Fr\'echet mean and its use in regression settings. 
Let $Y_i \ (i=1,\ldots,n)$ be observed data in a complete metric space $(\mathcal{U},d)$. 
The sample Fr\'echet mean is defined as 
$$
\bar{Y}= \mathop{\rm argmin}_{u\in\mathcal{U}}\sum_{i=1}^n d(Y_i,u)^2,
$$
where $d(\cdot, \cdot)$ is a distance. 
The existence of $\bar{Y}$ is always guaranteed, although uniqueness depends on the curvature properties of the metric space (e.g., Hilbert spaces or non-positively curved spaces ensure uniqueness). 
When the associated (Euclidean) covariate $X_i$ is available, the global Fr\'echet regression function \citep{petersen2019frechet} can be obtained by
\begin{equation}\label{eq:GF}
m(x)=\mathop{\rm argmin}_{u\in\mathcal{U}} \sum_{i=1}^n g(X_i,x)d(Y_i, u)^2,
\end{equation}
where 
\begin{equation}\label{eq:G(X)}
g(X_i,x)=1+(X_i-\mu_X)\Sigma_X^{-1}(x-\mu_X)
\end{equation}
with sample mean $\mu_X$ and covariance matrix $\Sigma_X$. 
The weight function $g(X_i,x)$ corresponds to the leverage structure in global least squares regression, so that all observations contribute to the estimate of $m(x)$ for any $x$. 
This global borrowing of information stabilizes estimation, especially with small sample sizes, but also makes the method less adaptive to local nonlinear structures.

A potential problem of the regression model (\ref{eq:GF}) is that it could be influenced by outlying objects.
A random object $Y_i$ is considered to be an outlier with respect to a given $x_i$ if the metric distance $d(Y_i, Y(x))$ is significantly large for $x$ in a neighborhood of $x_i$, where $Y(x)$ is a random object given $x$.
Such observation would have a large value of $g(X_i,x)d(Y_i, u)^2$ given the regression function $u$. 
Because the global regression uses all observations for any $x$, the effect of such outlying objects can propagate across the entire covariate space, leading to a biased estimate of $m(x)$ even at points far from $x_i$.

\subsection{Robust Fr\'echet regression with weight regularization}

To robustify the objective function (\ref{eq:GF}), we propose the following weighted loss formulation:  
\begin{equation*}
\begin{aligned}  
\label{weight-penalty-function}  
 & L(u, w; x) = \sum_{i=1}^n W_i g(X_i, x) d(Y_i, u)^2,
\end{aligned}  
\end{equation*} 
where  $1\ge W_i \ge 0$ is a weight parameter. 
Here $w = \{W_1, \ldots, W_n\}$ represents a set of weights, and the weight $W_i$ plays a critical role in determining the contribution of each observation to the loss function. 
Specifically, when $W_i = 1$, the corresponding observation $Y_i$ is fully utilized in the estimation process, whereas if $W_i = 0$, the information from $Y_i$ is entirely excluded.  
Initially, the weights $W_i$ should adaptively reflect the outlyingness of each observation, such that $W_i = 1$ for genuine (non-outlying) observations and $W_i = 0$ for outliers. Since the classification of observations as outlying or non-outlying is unknown a priori, $W_i$ is treated as an unknown parameter to be jointly estimated alongside the regression function $u$.

While $w$ is a high-dimensional parameter, we can assume sparsity for $w$ in the sense that most elements in $w$ are $1$, indicating that most observations are genuine (non-outlying) observations. 
Hence, in the estimation of $w$, we introduce a regularization term, where a similar approach is typically adopted in the estimation of Shift in the robust regression \citep{she2011outlier}. 
Specifically, we employ the Elastic net penalty \citep{zou2005regularization} for $1-W_i$. 
This penalty simultaneously enforces sparsity and smoothness in the estimation of $W_i$, facilitating effective differentiation between outliers and non-outlying observations.  
We therefore define the robust Fr\'echet regression function with the following objective function: 
\begin{equation}  
\label{weight-penalty-function}  
Q(u, w) 
= \sum_{i = 1}^n \Big\{
W_i g(X_i, x) d^2(Y_i, u) + \lambda |1-W_i| + \gamma (1-W_i)^2 
\Big\}, 
\end{equation} 
where $\lambda$ and $\gamma$ are tuning parameters.
Then, the regression function and weight parameter can be obtained as $(\hat{u},\hat{w})={\rm argmin}_{u\in\mathcal{U},w\in [0,1]^n} Q(u,w)$.
This optimization problem can be easily solved by an iterative algorithm described in Section~\ref{sec:algorithm}.
Given the regression function $u$, the optimal weight minimizing (\ref{weight-penalty-function}) can be obtained as follows: 

\begin{prp}
The optimal weight $(\Wt_1(u),\ldots,\Wt_n(u))={\rm argmin}_{w\in [0,1]^n} Q(u,w)$ is obtained as \begin{equation}\label{eq:weight-function}
\Wt_i(u) =
\begin{cases}
1, & g(X_i, x) d^2(Y_i, u)\in [-\infty, \lambda]\\
1 - \dfrac{1}{2\gamma}\bigl\{ g(X_i, x)d^2(Y_i, u) - \lambda \bigr\}, & g(X_i, x) d^2(Y_i, u)\in (\lambda, \lambda+2\gamma)\\
0, & g(X_i, x) d^2(Y_i, u) \in [\lambda + 2\gamma, \infty)
\end{cases}
\end{equation}
\end{prp}

The derivation is given in the Appendix.
From the expression (\ref{eq:weight-function}), the role of $W_i$ is more evident. 
The tuning parameters $\lambda$ and $\gamma$ determine the threshold for the weighted distance $g(X_i, x) d^2(Y_i, u)$, and the corresponding observation is recognized as outlier (i.e. $\Wt_i(u)=0$) and completely eliminated from the objective function when the weighted distance is larger than $\lambda+2\gamma$.
In contrast, when the weighted distance is smaller than $\lambda$, the weight parameter is exactly $1$, leading to the use of full information of $Y_i$ in the estimation of $u$.

In Figure~\ref{Fig.main2}, we present the shape of the adaptive weight $\Wt_i(u)$ as a function of the weighted distance $g(X_i, x) d^2(Y_i, u)$ under four cases of $(\lambda,\gamma)$. 
The curve of $W_i$  decreases as $d^2(Y_i, u)$ increases and $\lambda >0$ and $\gamma >0$ as Figure \ref{Fig.main2} .
The tuning parameter \( \lambda \) establishes the threshold for \( W_i = 1 \), controlling the quantity of total normal values. Conversely, for a fixed value of \( \lambda \), the Tuning parameter \( \gamma \) determines the threshold for \( W_i = 0 \), which controls the quantity of outlier values. Although $W_i$ equals 1 whenever $g(X_i,x)<0$, the sub-Gaussian assumption imposed in the convergence analysis guarantees that $g(X_i,x)$ remains bounded with exponential tails. As the result, the product $g(X_i,x),d^2(Y_i,u)$ cannot reach excessively a large value.

% Figure 
\begin{figure}[htb!] 
\centering 
\includegraphics[width=\textwidth]{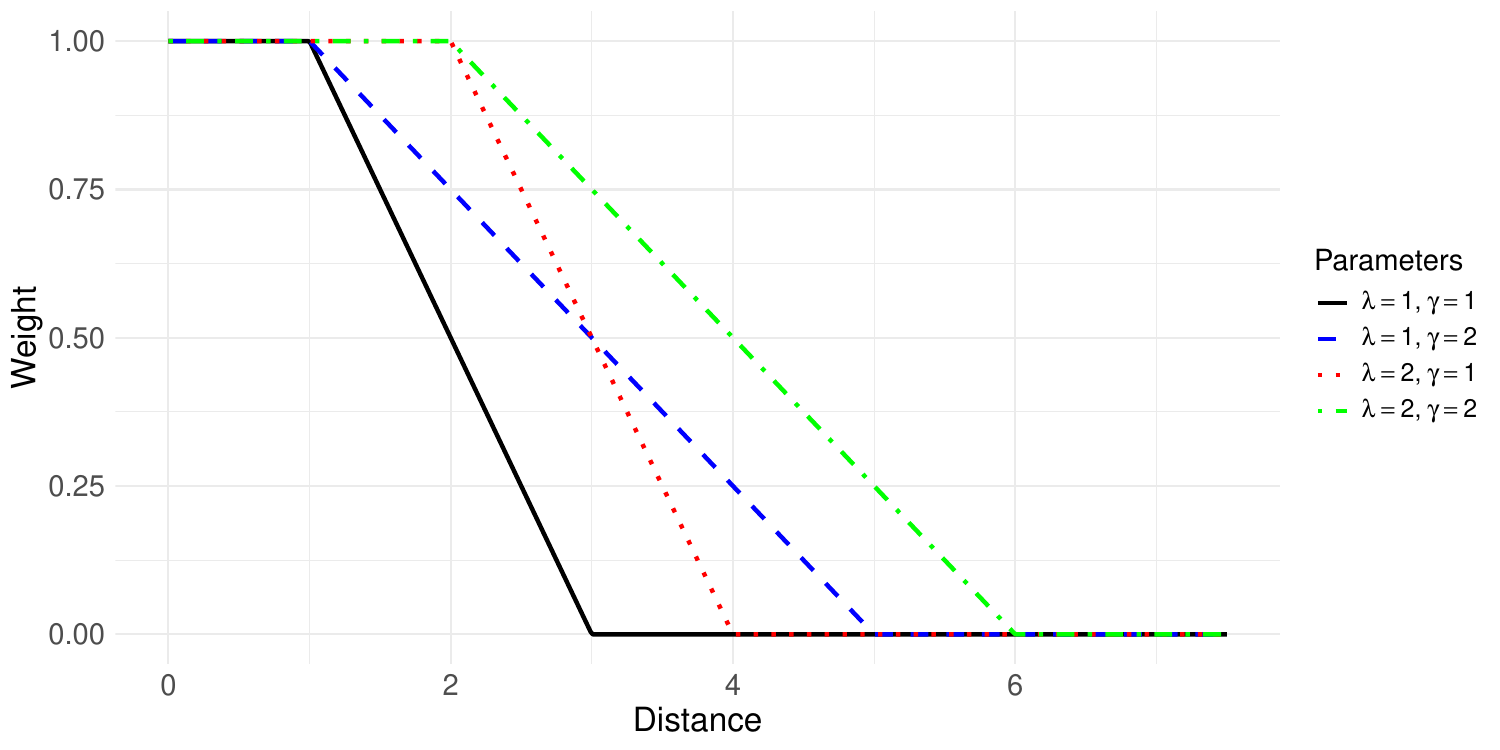} 
\caption{The adaptive weight function $\Wt_i(u)$ as a function of the weighted distance $g(X_i, x) d^2(Y_i, u)$ under four choices of $\lambda, \gamma\in \{1,2\}$.} 
\label{Fig.main2} 
\end{figure}

Using the adaptive weight function (\ref{eq:weight-function}), the profiled loss function for $u$ can be obtained as 
$$
\widetilde{Q}(u)\equiv \min_{w\in [0,1]^n}Q(u,w)=\sum_{i=1}^n 
\Wt_i(u) g(X_i, x) d^2(Y_i, u) +  B(u), 
$$
where $B(u)=\sum_{i=1}^n \lambda |1-\Wt_i(u)| + \gamma \{1-\Wt_i(u)\}^2$.
The first term of $\widetilde{Q}(u)$ can be regarded as a version of weighted least squares \citep{kiers1997weighted} for the global Fr\'echet regression with robust weight $\Wt_i(u)$.
Regarding the second term $B(u)$, we let $\omega$ be a proportion of outlying observations such that $n^{-1}\sum_{i=1}^n I\{\Wt_i(u_{\ast})<1\}=\omega \in (0,1)$ for the true regression function $u_{\ast}$. 
Then, it holds that $0\leq B(u_{\ast}) \leq \omega n$, indicating that the second term could be negligible when $\omega$ is relatively small (e.g. $\omega=0.05$).
Hence, the profiled objective function $\widetilde{Q}(u)$ is approximately equal to the outcome-dependent weighted loss function around the neighborhood of the true regression function $u_{\ast}$.
A notable feature of the joint minimization of $Q(u,w)$ is that the optimization can be efficiently performed through an iterative algorithm.

\subsection{On the penalty for weight parameters}\label{sec:penalty}
We here discuss the necessity of the Elastic net penalty in (\ref{weight-penalty-function}).
Following the regularized estimation of the shift value, one may consider the following $L_1$-penalized objective function:  
\begin{equation*}  
Q^{\dagger}(u, w) 
= \sum_{i = 1}^n \Big\{
W_i g(X_i, x) d^2(Y_i, u) + \lambda |1-W_i|
\Big\}, 
\end{equation*} 
instead of the Elastic net penalty given in (\ref{weight-penalty-function}).
Note that the objective function $Q^{\dagger}(u, w)$ is equivalent to (\ref{weight-penalty-function}) with $\gamma=0$ (without quadratic term).

Given $u$, the function $W_i g(X_i, x) d^2(Y_i, u) + \lambda |1-W_i|$ is increasing on $W_{i}\geq 1$, and reduces to $W_i g(X_i, x) d^2(Y_i, u) + \lambda (1-W_i)$ on $W_{i}\leq 1$.
Then, the optimal weight parameter $W_i$ given $u$ is obtained as 
\begin{equation*}
\Wt_i^{\dagger}(u) =
\begin{cases}
1, & g(X_i, x) d^2(Y_i, u)\in (-\infty, \lambda)\\
[0,1], & g(X_i, x) d^2(Y_i, u)=\lambda\\
0, & g(X_i, x) d^2(Y_i, u) \in (\lambda, \infty).
\end{cases}
\end{equation*}
A main drawback of the above weight is that the value is not uniquely determined for some observations. Moreover, its non-uniqueness depends on the tuning parameter, which also makes the tuning parameter selection challenging.
This is because the objective function $Q^{\dagger}$ as a function of $W_i$ is not strictly convex.
Therefore our alternative is using the Elastic Net penalty used in our proposal, which gives the unique weight as given in (\ref{eq:weight-function}).

\subsection{Optimization algorithm and its convergence property}\label{sec:algorithm}
The objective function (\ref{weight-penalty-function}) can be easily optimized by iteratively updating $u$ and $w$. 
The pseudo-code is given in Algorithm~\ref{algo:iterative}.
Note that the updating step for $u(x)$ is equivalent to conducting the Fr\'echet regression with $W_i^{(s+1)}g(X_i, x)$ being the weight for the distance, which enables us to employ the existing algorithm for the Fr\'echet regression.
In particular, we will demonstrate that the updating step is obtained in an analytical way under network and distribution responses.

\begin{algorithm}\label{algo:iterative}
\caption{Robust global Fr\'echet regression with weight regularization}  
\begin{algorithmic}[1]  
    \STATE Compute initial function $u^{(0)}(x)$ of $u(x)$ via the standard Fr\'echet regression by minimizing $Q(u,w)$ with $W_i=1$ and set $s=0$

    \REPEAT  
        \STATE Given $u^{(s)}(x)$, update the weight as $W_i^{(s+1)} \leftarrow \Wt_i(u^{(s)}(x))$ from (\ref{weight-penalty-function}). 
        \STATE Given $W_1^{(s+1)},\ldots,W_n^{(s+1)}$, update the regression function as 
        $$
        u^{(s+1)}(x) \ \leftarrow\ \mathop{\rm argmin}_{u\in\mathcal{U}}\sum_{i=1}^n W_i^{(s+1)}g(X_i,x)d^2(y_i, u).
        $$
        \STATE Set $s \leftarrow s+1$
        
    \UNTIL{ $d(u^{(s+1)}(x), u^{(s)}(x))< \epsilon$ }     
%    \RETURN \( Q^{-1}(z) \) and \( W_i \)  
\end{algorithmic}  
\end{algorithm}

Owing to the quadratic penalty term, $(1-W_i)^2$, in the proposed loss function (\ref{weight-penalty-function}), the solution is uniquely determined as explained in Section~\ref{sec:penalty}, which leads the linear convergence property of Algorithm~1. 
We assume the following regularity conditions: 

\begin{itemize}
\item[(C1)]
There exists a constant, $D_u>0$, such that $d(Y_i,u)\leq D_u$ for all $i$ and $u\in\mathcal U$.

\item[(C2)]
There exists a constant, $D_g>0$, such that $|g(X_i, x)|\leq D_g$ for all $i$ and $x\in \mathcal{X}$.
%$1 < g(X_i, x)<\infty$ for all $x\in \mathcal{X}$.

\item[(C3)]
$w=(W_1,\ldots,W_n)\in \mathcal{W}=\{w\in [0,1]^n \mid \sum_{i=1}^n W_i\geq \xi n\}$ for some $\xi\in (0,1)$.

\item[(C4)]
There exists a constant $L_d>0$ such that $|d^2(Y_i,u_1)-d^2(Y_i,u_2)| \leq L_d d(u_1,u_2)$ for all $i$ and $u_1,u_2\in\mathcal{U}$.

\item[(C5)]
For $w\in \mathcal{W}$, define a map $\Phi(w)$ as 
$$
\Phi(w)=\mathop{\rm argmin}_{u\in\mathcal U} \sum_{i=1}^n W_i g(X_i, x) d^2(Y_i,u).
$$
There exists a constant $C_u>0$ such that $d(\Phi(w_1), \Phi(w_2)) \leq C_u \|w_1-w_2\|$ for all $w_1,w_2\in \mathcal{W}$.
\end{itemize}

The conditions (C1) and (C2) are finiteness of the metric space and weight values. 
The condition (C3) is related to the maximum number of outliers. 
If the number of samples meeting $\sup_{x\in\mathcal{X},u\in\mathcal{U}}g(X_{i},x)d^{2}(Y_{i},u) < \lambda$, regarded as genuine observations, is larger than $\xi n$ then (C3) holds from (\ref{eq:weight-function}).
The conditions (C4) and (C5) are the Lipschitz conditions for the function $d^2(Y_i,\cdot)$ and the updating function $\Phi(w)$ for $u$ given $w$. 
Then, we have the linear convergence property of Algorithm~1.

\begin{prp}
Under regularity conditions (C1)-(C5), define  $\rho = (2\gamma)^{-1}\sqrt{n} C_u L_d D_g$. Then $d(u^{(s)},u^{\ast})\leq \rho^s d(u^{(0)},u^{\ast})$, for the minimizer $u^{\ast}$, so that Algorithm~1 exhibits linear convergence if $\rho<1$. 
\end{prp}

\subsection{Selection of the tuning parameter}
There are two tuning parameters $\lambda$ and $\gamma$ in the objective function (\ref{weight-penalty-function}), which would significantly control the downweighting of outliers as in (\ref{eq:weight-function}).
Here, we propose a data-dependent method for selecting the tuning parameters. 
Let $\hat{W}_i(\lambda, \gamma)$ and $\hat{u}(X_i; \lambda, \gamma)$ be estimates minimizing (\ref{weight-penalty-function}) with fixed $(\lambda, \gamma)$.
Then, the square of ``residual" can be defined as $d^2\{Y_i,\hat{u}(X_i; \lambda, \gamma)\}$.
Based on this quantity, we employ the following Bayesian information criterion \citep{gao2016penalized}: 
\begin{equation}\label{eq:BIC}
{\rm BIC}(\lambda,\gamma) = n \log \left\{ \frac{\sum_{i=1}^{n} \hat{W}_i(\lambda, \gamma) d^2\{Y_i,\hat{u}(X_i; \lambda, \gamma)\}}{\sum_{i=1}^{n} \hat{W}_i(\lambda, \gamma) } \right\} + \hat{k}({\lambda,\gamma}) \{\log(n) + 1\},  
\end{equation}
where $\hat{k}(\lambda, \gamma) = \sum_{i=1}^n {\rm I}\{\hat{W}_i(\lambda, \gamma) < 1\}$ is the number of ``outliers" under the tuning parameter $(\lambda,\gamma)$.
A similar criterion was introduced in \cite{she2011outlier}.
The BIC formula (\ref{eq:BIC}) indicates a trade-off between the goodness of fit and the number of suspected outliers.
In fact, the first term in (\ref{eq:BIC}) measures the goodness of fit while the second term measures the model robustness.

The optimal tuning parameter can be defined as the minimizer of the criterion (\ref{eq:BIC}). 
However, according to \cite{she2011outlier}, when the selected values of $\lambda$ and $\gamma$ result in a huge number of estimated outliers, it is often observed that the discriminative power of BIC substantially deteriorates. Therefore, it is recommended to define the lower bounds of $\lambda$ and $\gamma$  as the values corresponding to when the number of outliers exceeds 30\% of the total sample size, and the upper bounds as the maximum values ensuring that all data points are classified as non-outliers. Within this bounded interval, parameter selection and model screening based on BIC should be conducted to enhance the robustness and accuracy of outlier detection.

%---------------------------------------%
%             Illustration              %
%---------------------------------------%
\section{Illustrative Models }\label{sec:models}

\subsection{Robust regression for network and matrix response with Frobenius metric}

When $Y_i$ is a matrix or network, one may use the Frobenius metric \citep{hitchin1997frobenius} defined as 
$d(L_1, L_2)=\sqrt{ \text{tr}\left[ (L_1-L_2)^{\top} (L_1-L_2) \right]}$ for some matrices $L_1$ and $L_2$. 
In this case, the updating step for $u$ given $w$ is equivalent to minimizing 
\begin{equation*}
\sum_{i =1}^n W_i g(X_i,x) \  \text{tr} \left\{ (Y_i-u)^{\top} (Y_i-u) \right\},
\end{equation*}
and the optimal $u$ can be obtained as a weighted average as follows: 
\begin{equation}\label{eq:matrix-update}
\Phi(w; x) = \frac{\sum^n_{i = 1} W_i g(X_i,x) Y_i}{\sum^n_{i = 1}W_i g(X_i,x)}.
\end{equation}
Hence, the updating steps for $u$ and $w$ in Algorithm~\ref{algo:iterative} can be expressed in closed forms, so that the optimization problem can be easily solved. 
We can show that the Frobenius metric and the updating function (\ref{eq:matrix-update}) satisfies the regularity conditions, (C4) and (C5), required in Theorem~1, where the details are given in the Appendix.

\subsection{Robust regression for distribution response with Wasserstein distance}

When $Y_i$ is a distribution, $L_2$-Wasserstein distance can be employed to quantify the differences between two distributions. 
Regarding the $L_2$-Wasserstein distance \citep[e.g.][]{panaretos2016amplitude,turner2014frechet} between two distribution functions $F_1, F_2$ can be defined as $d(F_1, F_2) =  \| F_1^{-1} - F_2^{-1} \|_{2}$, where $F_1^{-1}(z)$ and $F_2^{-1}(z)$ represent the quantile functions for $z \in [0,1]$ and $\|\cdot\|_2$ denotes $L^2$-norm.
Under the settings, the updating step for $u$ given $w$ is 
\begin{equation*}
\sum_{i = 1}^n W_i  g(X_i,x) \| F_i^{-1} - u \|_{2}^2,
\end{equation*}
where $F_i^{-1}$ is a quantile function induced from a distribution observation $Y_i$. The above optimization problem gives the closed-form expression for $u$ given by 
\begin{equation*}
\Phi(w; x, z)= \frac{\sum^n_{i = 1} W_i g(X_i,x) F_i^{-1}(z)}{\sum^n_{i = 1}W_ig(X_i,x) },
\label{distrubution-result}
\end{equation*}
which is a weighted average of the quantile functions. 
As in the matrix response, we can show that the regularity conditions, (C4) and (C5), are satisfied in the settings, where the details are given in the Appendix.

%---------------------------------------%
%         Numerical results             %
%---------------------------------------%
\section{Numerical Studies}\label{sec:example}

\subsection{Simulation experiment with matrix response}

We evaluate the numerical performance of the proposed robust Fr\'echet regression via simulation experiments under matrix response in the following two cases of data generating process.

\begin{itemize}
\item[(I)]
We generate a univariate covariate $X$ from the uniform distribution on the interval \([0,1]\), i.e.,
$X \sim U(0,1)$.
Let $Y$ be a $q\times q$ matrix whose diagonal elements are 1 and off-diagonal elements, $Y_{jk} \ (j\neq k)$, are generated from the beta distribution, $\text{Beta}(X, 1 - X)$. 
Note that $E[Y_{jk}]=X$ for $j\neq k$ and the true regression value at $X=x$ is $M_{\ast}(x)=x I_{q} + (1-x) J_{q}$, where $I_{q}$ is the $q \times q$ identity matrix and $J_{q}$ denotes the $q \times q$ matrix of all ones.

\item[(II)]
We generate $q$-dimensional covariate $X$ from the uniform distribution on $[0,1]^p$, and the response matrix $Y$ is generated via symmetric matrix variate normal distribution, following \cite{qiu2024random}.
The $(j,k)$-element of $Y$, denoted by $Y_{jk}$, is defined as $Y_{jk}=\exp\{0.2Z_{jk}+D_{jk}(X)\}$, where $Z_{jk}\sim N(0, 1)$ for $j=k$ and $Z_{jk}\sim N(0, 1/2)$ for $j\neq k$.
Here $D_{jk}(X)=1$ for $j=k$ and $D_{jk}(X)=U_{jk} \cos(4 \pi (\beta^T X))$ with $U_{jk}\sim U(0,1)$, where $\beta = (0.1,\, 0.2,\, 0.3,\, 0.4,\, 0.5,\, 0,\ldots, 0)^\top$.
\end{itemize}
In this experiment, we considered two cases for the sample size, $n\in \{50, 100\}$ and set $q=8$ in DGP (I) and $q=10$ in DGP (II). 
To simulate outlier scenarios, we randomly sampled 10\% and 20\%  of observations from the full dataset to form two subsets. For each selected observation, we introduced synthetic outliers by adding a fixed additive shift value of either 50 or 100 to every element of the corresponding matrix.

For the generated dataset, we applied the standard and the proposed robust Fr\'echt regression.
We evaluate the estimation results for a newly generated covariate $\tilde{x}_i$ and its corresponding target $M_{\ast}(\tilde{x}_i)$, computing the mean squared error defined as:
\begin{equation*}
{\rm MSE} = \frac{1}{n}\sum^n_{i = 1}
\text{tr}\left[ \big\{\widehat{M}(\tilde{x}_i)-M_{\ast}(\tilde{x}_i)\big\}^{\top} \big\{\widehat{M}(\tilde{x}_i)-M_{\ast}(\tilde{x}_i)\big\} \right],
\label{MSE-matrix}
\end{equation*}
where $\widehat{M}(x_i)$ is the estimated regression function.

Table~\ref{tab:matrix-sim} summarizes the average MSE values averaged over 100 Monte Carlo replications of the standard and robust Fr\'echet regression estimators under two scenarios of DGP and five contamination settings. 
The Monte Carlo standard errors are reported in parentheses.
Under DGP (I) without contamination, both estimators exhibit nearly identical MSEs, indicating that the robust modification preserves efficiency in the uncontaminated setting. 
As the contamination proportion increases, the MSE of the non-robust method rises sharply, whereas the robust Fr\'echet regression method displays only a mild increase. 
This tendency holds for both $n=50$ and $n=100$. 
The corresponding relative MSEs (standard over robust) exceed 20 in the highest contamination scenarios.
In DGP (II), the original Fr\'echet regression method has a modest advantage in the uncontaminated setting, while the original Fr\'echet regression exhibits a much deeper escalation of MSE as the contamination ratio increases than the robust Fr\'echet regression.
We further report the BIC-selected tuning parameters \(\lambda\) and \(\gamma\) value across data scenarios of DGP (I); the results are presented in the Table \ref{tab:lam-res-matrix}. We find that, as the contamination proportion and the shift value increase, the BIC-selected \(\lambda\) and \(\gamma\) also increase. These results suggest that \(\gamma\) also plays a critical role in downweighting the influence of heavily contaminated and high-bias observations.

In conclusion, these findings confirm that the robust Fr\'echet regression preserves efficiency in clean samples while offering substantial protection against contamination, with benefits increasing with both the contamination proportion and the magnitude of the shift value. The improvements are particularly striking for DGP (I), where the robust Fr\'echet regression method nearly eliminates the adverse effects of even severe contamination.

% Table (MSE)
\begin{table}[htb!]
\centering  
\begin{tabular}{ccccccccccccccc}
\hline
&& Proportion & 0 & 0.1 & 0.1 & 0.2 & 0.2 \\
DGP &$n$ & Shift  & - & 50 & 100 & 50 & 100\\
\hline
(I) &50 & Standard & 0.48 {\scriptsize (0.01)} & 61.3 {\scriptsize (2.6)} & 120.3 {\scriptsize (4.9)} & 100.6 {\scriptsize (1.7)} & 199.3 {\scriptsize (4.8)} \\
&& Robust   & 0.48 {\scriptsize (0.01)} & 1.7 {\scriptsize (0.2)} & 2.0 {\scriptsize (0.2)} & 6.9 {\scriptsize (0.5)} & 9.6 {\scriptsize (1.0)} \\
\hline
(I) &100 & Standard & 0.31 {\scriptsize (0.00)} & 52.3 {\scriptsize (1.3)} & 103.5 {\scriptsize (2.5)} & 101.5 {\scriptsize (2.3)} & 202.0 {\scriptsize (4.6)} \\
&& Robust   & 0.31 {\scriptsize (0.00)} & 1.6 {\scriptsize (0.2)} & 2.5 {\scriptsize (0.3)} & 6.6 {\scriptsize (0.9)} & 10.2 {\scriptsize (1.2)} \\
\hline
(II) &50 & Standard & 17.1 {\scriptsize (4.9)} & 100.6 {\scriptsize (4.6)} & 152.0 {\scriptsize (6.6)} & 146.8 {\scriptsize (5.5)} & 147.1 {\scriptsize (5.2)} \\
&& Robust   & 25.6 {\scriptsize (1.8)} & 27.6 {\scriptsize (1.5)} & 34.2 {\scriptsize (2.1)} & 37.1 {\scriptsize (2.2)} &  45.84 {\scriptsize (2.6)} \\
\hline
(II) & 100 & Standard & 33.4 {\scriptsize (2.3)} & 94.9 {\scriptsize (3.3)} & 140.7 {\scriptsize (4.2)} & 146.2 {\scriptsize (4.8)} & 232.9 {\scriptsize (8.9)} \\
&& Robust  & 20.4 {\scriptsize (1.4)} & 25.2 {\scriptsize (14.7)} & 41.8 {\scriptsize (2.3)} & 29.4 {\scriptsize (2.3)} & 50.1 {\scriptsize (2.9)} \\
\hline
\end{tabular}
\caption{
The averaged MSE of the standard and robust Fr\'echet regression under matrix response with 8 or 10 dimensions, sample sizes of $50$ and $100$, five scenarios of contamination and two cases of data generating process (DGP).
The values are based on 100 replications, and the estimated Monte Carlo errors are given in parentheses.  
}
\label{tab:matrix-sim}
\end{table}

% Table (tuning parameter)
\begin{table}[htb!]
\centering
\begin{tabular}{cccccccccccccccccccc}
\hline
 & & \multicolumn{2}{c}{(0, -)} & \multicolumn{2}{c}{(0.1, 50)} & \multicolumn{2}{c}{(0.1, 100)} & \multicolumn{2}{c}{(0.2, 50)} & \multicolumn{2}{c}{(0.2, 100)} \\
 $n$ && $\lambda$ & $\gamma$ & $\lambda$ & $\gamma$& $\lambda$ & $\gamma$& $\lambda$ & $\gamma$& $\lambda$ & $\gamma$\\
\hline
 50 & & 0.37 & 0.00 & 0.74 & 0.00 & 1.39 & 0.20 & 2.02 & 0.28 & 2.95 & 1.27 \\
 100 & & 0.59 & 0.00  & 1.63 & 0.52 & 1.89 & 1.39 & 2.92 & 0.66 & 3.41 & 1.26 \\
\hline
\end{tabular}%
\caption{The value of $\lambda$ and $\gamma$ for matrix response under various data configurations (sample size $n$ and other parameter settings). Values are multiplied by $10^2$.}
\label{tab:lam-res-matrix}
\end{table}

\subsection{Demonstration using the New York Yellow Taxi network data}

We next demonstrate the robust Fr\'echet regression with network response through the dataset from the TLC Trip Record Data provided by the New York City Taxi \& Limousine Commission. 
It comprises detailed trip records, including 143 days of data for yellow taxis operating within New York State. The data includes information on pickup and drop-off data and times, pickup and drop-off locations, trip distances, itemized fare components, fare structures, and payment methods.
All data available at \url{https://www1.nyc.gov/site/tlc/about/tlc-tr ip-record-data.page}.

We focused on investigating the transportation network's dependency, constructed from taxi trip records, on new COVID-19 cases and the weekend indicator. 
To simulate the case of outliers, we randomly select 10\% of the data in the transportation network and add a residual of 100 to each element within the contaminated transportation network data.
The model optimization process employed the Bayesian Information Criterion (BIC) as the primary parameter selection metric. For validation purposes, we adopted a leave-one-out validation strategy wherein a single observation from the uncontaminated (normal) dataset was randomly designated as the test sample, while the remaining observations, including both clean and contaminated data points, formed the training set. Following model estimation on the training data, the model's predictive performance was evaluated by computing the Mean Square Error (MSE) on the held-out test sample, as defined in Equation \ref{MSE-matrix}. The results are as follows \ref{tb:taxi_table}.

The absolute error heat maps in Figure \ref{Fig.main2} reveal that the robust method consistently yields low prediction errors across the network, with only a few localized regions showing moderate deviations. In contrast, the network regression method exhibits concentrated zones of higher error, while the non-robust method suffers from widespread large deviations, as indicated by extensive high-intensity red areas.
The quantitative comparison in Table \ref{tb:taxi_table} further confirms these observations. The robust method achieves a substantially lower mean squared error compared with the network regression method and the non-robust method. 
The reduction in both mean error magnitude and variability demonstrates the robustness of the proposed approach in reducing the influence of outliers.
Overall, the results provide strong evidence that the robust method demonstrates superior predictive accuracy and stability compared with contaminated, real-world conditions data.

\begin{table}[htb!]
\label{tb:taxi_table}
\centering
\begin{tabular}{ccccccc}
\hline
method& & non-robust   &network regression  & robust    \\ \hline
MSE   & & 4140 {\scriptsize (2147)}  &  3522 {\scriptsize (1551)}      & 872 {\scriptsize (385)}    \\ \hline
\end{tabular}
\caption{The leave-one-out MSE to measure the New York Yellow Taxi System.}
\end{table}

\begin{figure}[htb!] 
\centering 
\includegraphics[width=1\textwidth]{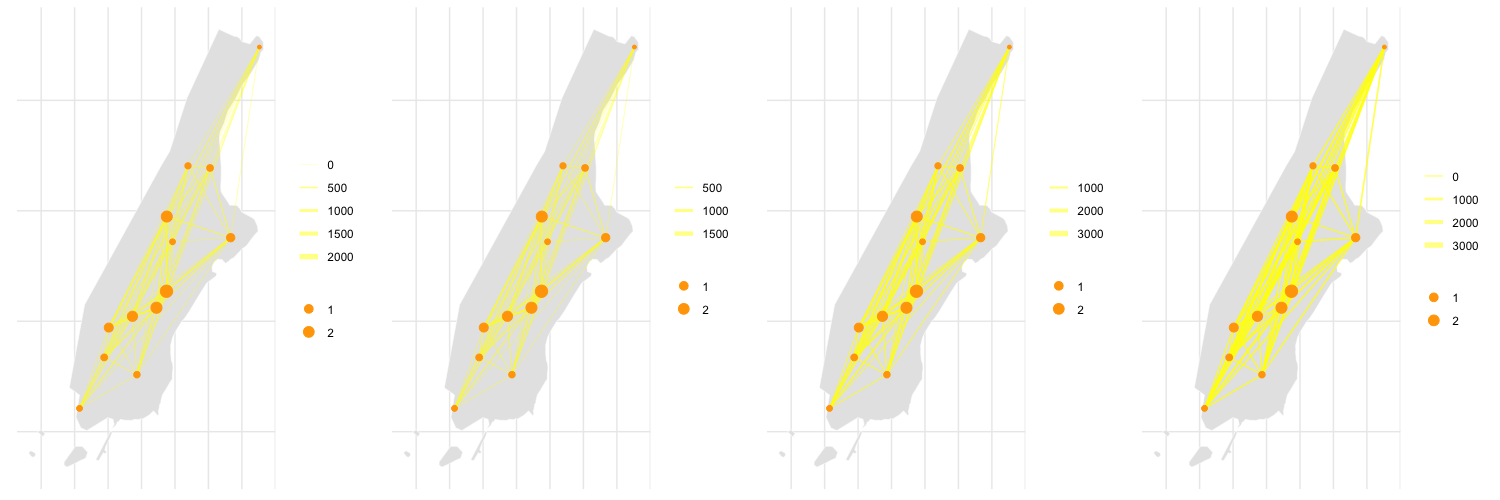} 
\caption{True networks(first), robust fitted networks(second), network regression method proposed by \cite{zhou2022network}(third), and non-robust fitted networks (fourth) on May 16, 2020, corresponds to the day when the number of new COVID-19 cases was 134.} 
\label{Fig.main1} 
\end{figure}

\begin{figure}[htb!] 
\centering 
\includegraphics[width=0.8\textwidth]{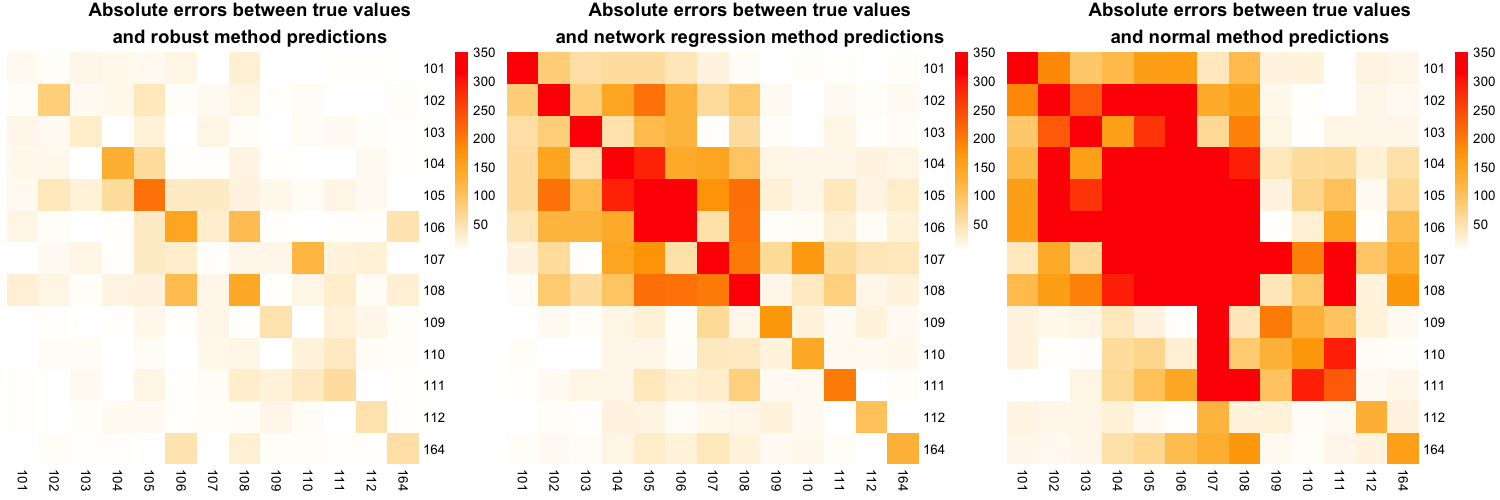} 
\caption{The absolute error heat map between the true network value, the network regression method proposed by \cite{zhou2022network}, and the predicted network value on May 16, 2020, corresponds to the day when the number of new COVID-19 cases was 134.} 
\label{Fig.main2} 
\end{figure}

\subsection{Simulation experiment with distribution response}

To assess the performance in the distribution response. 
For each observation \( i = 1, \ldots, n \), the covariate \( X_i \) is independently drawn from a uniform distribution on \([0,1]\). 
Conditional on \( X_i \), the true mean parameter \( \mu_i \) is generated from the normal distribution, $\mu_i |X_i \sim \mathcal{N}(\mu_0 + \beta X_i, v_1)$.
Also, the true standard deviation parameter \( \sigma \) is generated from a gamma distribution, $\sigma_i | X_i \sim {\rm Ga}(\alpha_i, \lambda_i)$, where $\alpha_i=(\sigma_0 + \gamma X_i)^2/v_2$ is a shape parameter and $\lambda_i=v_2/(\sigma_0 + \gamma X_i)$ is a rate parameter. 
For a fixed set of quantile levels \( \{ z_j \} \) as an equally spaced sequence starting from 0.1 to 0.9 with an increment of 0.01, expressed as $z_j = 0.1 + 0.01 \times (j - 1)$ for $j = 1, 2, \ldots, 81$.  
The response variable \( Y_{ij} \) at quantile \( z_j \) is generated via the quantile function of the standard normal distribution as $Y_{ij} = \mu_i + \sigma_i \Phi^{-1}(z_j)$, where \( \Phi(\cdot) \) denotes the standard normal distribution function.
To account for contaminated data, we randomly selected a predetermined number of samples from the observations and introduced a constant shift value to the corresponding response variable at each. Regarding generating contaminated data, we adopted the same strategy as with the matrix response. We randomly select a 10\% or 20\% dataset and add a 50 or 100 shift in each element of the corresponding distribution observation.

For the simulated data, we applied the standard and robust Fr\'echet regression.
We then measure the estimation accuracy via the mean integrated squared error (MISE) for a newly generated observation, defined as:
\begin{equation*}
{\rm MISE}= \frac{1}{n}\sum^n_{i = 1}  \int \big\{\widehat{F}_i^{-1}(z) - F_i^{-1}(z) \big\}^2 dz,
\end{equation*}
where $F_i^{-1}(z)$ is the true quantile function. 
The above integral is approximated by 81 grid points of quantile levels. 
We constructed the candidate set of $\lambda$ values as follows. First, we generated an equally spaced sequence $\{x_i\}_{i=1}^{20}$ over the interval $[10^{-7}, 1]$. We then mapped this grid to the $\lambda$ scale via $\lambda_{\max}\, x_i^{0.8}$, where $\lambda_{\max}$ denotes the largest $\lambda$ for which no observations are flagged as outliers (i.e., all points are classified as non-anomalous). Because larger values of $\lambda$ tend to increase the number of detected anomalies in our setting, we employed the exponent $0.8$ to induce a denser grid near smaller effective $\lambda$ values, thereby enabling a finer search in the more sensitive region of the parameter space.

Table \ref{tab:dist-sim} summarizes average MISE values averaged over 100 Monte Carlo replications with Monte Carlo standard errors (in parentheses) for the original and robust Fr\'echet regressions under distribution responses with $n=50$ and $n=100$ across two contamination scenarios. In the absence of contamination, both methods yield comparable MISEs. As the contamination proportion increases,  MISEs rise for both methods, but the increase is markedly more pronounced for the original Fr\'echet regression, especially under large contamination and shifts. 
For example, at $n=50$ with a contamination proportion of $0.2$ and shift of $100$, the MISE of the robust Fr\'echet regression is approximately a quarter of that of the original Fr\'echet regression. 
Similar to $n=100$, underscoring the superior stability of the robust Fr\'echet regression in the presence of outliers. The \(\lambda\) and \(\gamma\) results for the distribution response are also reported in Table \ref{tab:lam-res-distrubution}. Consistent with the matrix-response case, \(\gamma\) and \(\lambda\) remain large when greater shift value and higher contamination proportions.

\begin{table}[htb!]
\centering  
\begin{tabular}{ccccccccccccccc}
\hline
& Proportion & 0 & 0.1 & 0.1 & 0.2 & 0.2 \\
$n$ & Shift  & - & 50 & 100 & 50 & 100\\
\hline
50 & Standard Fr\'echet & 37.9 {\scriptsize (2.5)} & 67.0 {\scriptsize (2.1)} & 102.9 {\scriptsize (1.4)} & 104.2 {\scriptsize (1.6)} & 187.1 {\scriptsize (1.4)} \\
   & Robust Fr\'echet   & 38.3 {\scriptsize (2.5)} & 42.5 {\scriptsize (2.9)} & 41.5 {\scriptsize (2.8)} & 47.4 {\scriptsize (3.1)} & 47.4 {\scriptsize (3.0)} \\
\hline
100 & Standard Fr\'echet & 43.1 {\scriptsize (2.9)} & 65.8 {\scriptsize (2.1)} & 103.5 {\scriptsize (1.6)} & 100.0 {\scriptsize (1.4)} & 185.4 {\scriptsize (1.0)} \\
    & Robust Fr\'echet   & 42.0 {\scriptsize (2.9)} & 44.5 {\scriptsize (3.0)} & 43.7 {\scriptsize (2.9)} & 37.7 {\scriptsize (3.0)} & 37.9 {\scriptsize (3.0)} \\
\hline
\end{tabular}
\caption{
The mean integrated squared errors of the standard and robust Fr\'echet regression under distribution response with sample sizes of $50$ and $100$, and five scenarios of contamination. 
The values are based on 100 replications, and the estimated Monte Carlo errors are given in parenthesis.  
}
\label{tab:dist-sim}
\end{table}

\begin{table}[htb!]
\centering
\begin{tabular}{cccccccccccccccccccc}
\hline
 & & \multicolumn{2}{c}{(0, -)} & \multicolumn{2}{c}{(0.1, 50)} & \multicolumn{2}{c}{(0.1, 100)} & \multicolumn{2}{c}{(0.2, 50)} & \multicolumn{2}{c}{(0.2, 100)} \\
 $n$ && $\lambda$ & $\gamma$ & $\lambda$ & $\gamma$& $\lambda$ & $\gamma$& $\lambda$ & $\gamma$& $\lambda$ & $\gamma$\\
\hline
 50 & & 2.42 & 0.03 & 3.61 & 0.56 & 5.24 & 2.63 & 4.35 & 1.68 & 6.66 & 5.45 \\
 100 & & 2.89 & 0.59  & 4.80 & 0.61 & 5.80 & 3.62 & 6.23 & 5.66 & 6.52 & 6.97 \\
\hline
\end{tabular}%
\caption{The value of $\lambda$ and $\gamma$ for distribution response under various data configurations (sample size $n$ and other parameter settings). 
Values are multiplied by $10^4$.}
\label{tab:lam-res-distrubution}
\end{table}

\subsection{Illustration of distribution response with mortality Data}
Many studies and analyses have been motivated by a desire to understand human longevity. Of particular interest is the evolution of the distributions of age-at-death over calendar time. This database includes yearly mortality and population data for 37 countries that are available at \url{www.mortality.org}. As an initial example, we consider the data for Luxembourg, which has mortality data available for the years 1960–2009.
We employ an identical methodology for constructing the independent variables. The global Fr\'echet regression is fitted using the calendar year as the predictor variable for the quadratic model $(X_i = (t_i, t^2_i)^T)$. where $t_i = i + 1959$, $i = 1, \dots, 50$.

For the dataset, the proposed robust Fr\'echet regression was compared with the conventional non-robust approach under a quadratic model specification with calendar year as the predictor. Model performance was evaluated using leave-one-out MISE with Monte Carlo standard errors, which are reported in Table~\ref{tab:mortality}. 
The robust estimator achieved a substantially lower MISE compared to the non-robust method, indicating improved predictive accuracy and resistance to the influence of potential outlying observations in the mortality data.

\begin{table}[htb!]
\label{tab:mortality}
\centering
\begin{tabular}{lll}
\hline
 method   & robust          & non-robust  \\ \hline
MISE & 3.57 {\scriptsize (1.39)} & 6.56 {\scriptsize (1.56)} \\ \hline
\end{tabular}
\caption{Leave-one-out MISE of robust and non-robust version of the quadratic global Fr\'echet regression applied to Luxembourg mortality data, where the Monte Carlo standard errors are present in the parentheses.}
\end{table}

%---------------------------------------%
%         Concluding Remarks            %
%---------------------------------------%
\section{Concluding Remarks}
In this study, we base our work on the concept of Fr\'echet regression to develop a robust local Fr\'echet regression framework. We incorporate observation weight parameters into the original objective function of Fr\'echet regression. Since we need to downweight abnormal observations, we apply an Elastic Net penalty to $1-W_i$, thereby automatically controlling model robustness. To search for the best hyperparameter of penalty, we propose a data-driven tuning strategy based on the BIC. We demonstrate that under certain conditions, the proposed method is linear convergence. At least, we conduct comprehensive simulation studies to evaluate the proposed method both in the matrix space and distribution space. Additionally, real data analyses are performed for each case. The results consistently demonstrate that, compared with traditional models, our method exhibits superior robustness.

However, our method still has certain limitations. Specifically, it only assesses the overall outliers for each observation as a whole. It does not allow for the evaluation of the outliers of individual components within each observation. For instance, in the case of a matrix response, the anomaly of a single component may lead to the entire matrix being identified as an outlier. Nevertheless, in the estimation process, the presence of other normal components can help decrease the variance of the model. Therefore, extensions addressing this limitation will be considered in our future work.

%---------------------------------------%
%          Acknowledgement              %
%---------------------------------------%
\section*{Acknowledgement}
This work is partially supported by JSPS KAKENHI Grant Numbers 24K21420 and 25H00546.

\appendix
%---------------------------------------%
%             Appendix                  %
%---------------------------------------%
\begin{center}
{\large {\bf Appendix}}
\end{center}

\section{Proof of Proposition~1}

For notational simplicity, we let $r_i(u) = g(X_i, x)\, d^2(Y_i, u)$. Then, the adaptive weight $\widetilde{W}_i(u)$ is the minimizer of
\begin{equation*}
r_i(u) W_i + \lambda |1 - W_i| + \gamma (1 - W_i)^2
\end{equation*}
under $1\ge W_i \ge 0$, the objective function is strictly convex and can be rewritten as $\gamma (W_i - \bar{w}_i)^2 - \gamma \bar{w}_i^2 + \lambda + \gamma$, where $\bar{w}_i = 1 - (r_i(u) - \lambda)/2\gamma$.
We consider the case $\bar w_i \le 0$, i.e.,
$r_i(u) \ge \lambda + 2\gamma$, for which the minimizer achieve at the boundary $W_i = 0$. We then consider the case $0 < \bar w_i < 1$, equivalently, $\lambda < r_i(u) \le \lambda + 2\gamma$,
where the minimizer is in the interior of the feasible region, and hence
$\widetilde W_i(u) = \bar w_i$. Last,  for the case $\bar{w}_i \geq 1$ or equivalently $r_i(u) \leq \lambda$, the unconstrained minimum is greater than or equal to 1. The constrained minimizer is the projection of $\bar{w}_i$ onto the feasible region, resulting in $\widetilde{W}_i(u) = 1$.

\section{Proof of Proposition~2}

Let $\widetilde{w}(u)=(\widetilde{W}_1(u),\ldots,\widetilde{W}_n(u))$, where $\widetilde{W}_i(u)$ is defined in (\ref{eq:weight-function}).
We then define a mapping $T(u)=\Phi(\widetilde{w}(u))$ representing the one-step updating process of Algorithm~1.
We will show that $T(u)$ is a contraction on $\mathcal{U}$, under which the sequence $u_{(s+1)}=T(u^{(s)})$ linearly converges to a fixed point from the Banach fixed-point theorem.
For $r_i(u) = {g(X_i, x) d^2(Y_i, u)}$, it follows from (C2) that 
$$
|r_i(u_1)-r_i(u_2)|
= |g(X_i,x)|\cdot |d^2(Y_i,u_1)-d^2(Y_i,u_2)| \leq |g(X_i,x)| L_d d(u_1,u_2),
$$
for all $i$ and $u_1,u_2\in\mathcal U$.
Further, using the form of $\widetilde W_i(u)$ given in (\ref{eq:weight-function}), we have
$$
|\widetilde W_i(u_1)-\widetilde W_i(u_2)|
\leq \frac{1}{2\gamma}|r_i(u_1)-r_i(u_2)|
\leq \frac{1}{2\gamma} |g(X_i,x)| L_d d(u_1,u_2).
$$
Hence, it holds that $\|\widetilde{w}(u_1)-\widetilde{w}(u_2)\| \leq (2\gamma)^{-1}\sqrt{n}D_g L_d  d(u_1,u_2)$. 
Under (C5), it holds that 
\begin{align*}
d(T(u_1),T(u_2))
&= d(\Phi(\widetilde{w}(u_1)),\Phi(\widetilde{w}(u_2)))
\leq C_u\,\|\widetilde{w}(u_1)-\widetilde{w}(u_2)\|\\
&\leq \rho d(u_1,u_2),
\end{align*}
where $\rho=\sqrt{n}C_u D_g L_d /2\gamma$. 
When $\rho<1$, $T$ is a contraction mapping, which completes the proof.

\section{Regularity Conditions for Matrix and Distribution Response Models}

We check (C4) and (C5) for specific models described in Section~\ref{sec:models} and validate the condition, $\rho<1$, given in Proposition~2.
To this end, we also assume that $X_1,\ldots,X_n$ are independent and identically distributed random vectors, having mean $\mu_X$ and variance-covariance matrix $\Sigma_X$. 
We also assume that $X_i$ is sub-Gaussian.

\subsection{Matrix response with Frobenius norm}\label{sec:proof-matrix}
From the triangular inequality, it holds that 
\begin{align*}
\big|d^2(Y_i,u_1)-d^2(Y_i,u_2)\big|
&=\big|\big(\|u_1-Y_i\|_F-\|u_2-Y_i\|_F\big)
        \big(\|u_1-Y_i\|_F+\|u_2-Y_i\|_F\big)\big|\\
&\leq (\|u_1-Y_i\|_F+\|u_2-Y_i\|_F) d(u_1,u_2) \leq 2D_u d(u_1,u_2),
\end{align*}
whereby (C3) is satisfied with $L_d=2D_u$. 
Furthermore, we define $A(w)=\sum_{i=1}^n W_i g(X_i,x)Y_i$ and $S(w)=\sum_{i=1}^nW_i g(X_i,x)$, so that the updating function (\ref{eq:matrix-update}) can be expressed as $\Phi(w)=A(w)/S(w)$. 
Then, it holds that 
\begin{equation}\label{eq:Phi-diff}
\Phi(w_1) - \Phi(w_2) = \frac{A(w_1)-A(w_2)}{S(w_1)} + \frac{A(w_2)\{S(w_2)-S(w_1)\}}{S(w_1)S(w_2)}.
\end{equation}
Since $A(w_1)-A(w_2)=\sum_{i=1}^n (W_{i2}-W_{i1})g(X_i,x)Y_i$, it follows from the Cauchy-Schwartz inequality that  
\begin{align*}
\|A(w_1)-A(w_2)\|_F 
&\leq \|w_1-w_2\| \Big\{\sum_{i=1}^n \|g(X_i,x) Y_i\|_F^2\Big\}^{1/2}
\leq \sqrt{n}\|w_1-w_2\| D_{g} D_u,
\end{align*}
where $D_g=\max_{i=1,\ldots,n} |g(X_i,x)|$.
We also note that 
$$
\|A(w_2)\|_F \leq  \sum_{i=1}^n W_i |g(X_i,x)|\cdot\|Y_i\|_F 
\leq  n D_gD_u,
$$
and $|S(w_2)-S(w_1)|\leq \sqrt{n} D_g \|w_2-w_1\|$ from the Cauchy-Schwartz inequality.
To evaluate the behavior of $S(w)$ for some $w\in \mathcal{W}$, we note that $S(w)=\sum_{i=1}^n W_i +\sum_{i=1}^n W_ia_i$, where $a_i= (x-\mu_X)^\top\Sigma_X^{-1} (X_i-\mu_X)$ with $E[a_i]=0$ is a sub-Gaussian random variable with parameter denoted by $\sigma^2(x)$.
Since $\sum_{i=1}^n W_ia_i$ is also sub-Gaussian with parameter $\sigma^2(x)\sum_{i=1}^n W_i^2$, it follows from (C3) and the Hoeffding inequality that 
\begin{align*}
P\left( S(w)\geq \xi n -\epsilon \right)
\geq  1 - 2\exp\left(-\frac{\epsilon^2}{2\sigma^2(x)\sum_{i=1}^n W_i^2}\right) 
\geq 1 - 2\exp\left(-\frac{\epsilon^2}{2n\sigma^2(x)}\right).
\end{align*}
Given that $g_i = 1 + a_i$ and the noise term $a_i$ is sub-Gaussian with mean zero, the event $g_i < 0$ corresponds to $a_i < -1$, whose probability is exponentially small by standard sub-Gaussian tail bounds. Furthermore, on the event $g_i < 0$, the $|g_i|$ concentrated to zero. Thus, these observations can be asymptotically ignored.
Then, with probability approaching to 1, it holds that 
\begin{align}\label{eq:Phi-diff}
d(\Phi(w_1), \Phi(w_2))=\|\Phi(w_1) - \Phi(w_2)\|_F \leq \underbrace{ \left( \frac{n^{\frac{1}{2}}D_{g} D_u}{\xi n -\epsilon} +  \frac{n^{\frac{3}{2}}D^2_{g} D_u}{(\xi n -\epsilon)^2} \right)}_{C_u} \|w_2-w_1\|,
\end{align}
which gives (C4).

Furthermore, it hold that 
$$
\rho=\frac{\sqrt{n}C_u L_d D_g}{2\gamma}
=\frac{L_dD_g^2 D_u}{2\gamma}\left(\frac{n}{\xi n -\epsilon} + \frac{n^2 D_g}{(\xi n -\epsilon)^2}\right) \to \frac{L_dD_g^2 D_u (\xi + D_g)}{2\gamma \xi^2},
$$
as $n\to\infty$, whereby $\rho<1$ is established when $\gamma$ (tuning parameter in the weight regularization) and $\xi$ (related to the number of non-outliers) are large.

\subsection{Distribution response with $L_2$-Wasserstein distance}
According to the definition of $L_2$-Wasserstein distance, we obtain
\begin{align*}
&|d^2(Y_i,u_1) - d^2(Y_i,u_2)|\\ 
\ \ \ &= \bigl(\|u - F^{-1}_i(z)\|_2 - \|u - F^{-1}_i(z)\|_2\bigr) 
\bigl(\|u - F^{-1}_i(z)\|_2 + \|u - F^{-1}_i(z)\|_2\bigr)\\
&\le 2D_u d(u_1,u_2),
\end{align*}
whereby (C3) is satisfied with $L_d = 2 D_u$. 
Similarly to Section~\ref{sec:proof-matrix}, we obtain (\ref{eq:Phi-diff}) with probability approaching to 1 and $\rho<1$ is satisfied when $\gamma$ and $\delta$ are large.

\vspace{1cm}
%   Reference
\bibliographystyle{chicago}
\bibliography{ref}

\end{document}